\documentclass{ws-ijmpb}
\usepackage{graphicx,rotating,subfigure,amsmath,amsfonts,amssymb,delarray}
\usepackage[super,sort]{cite}

\newcommand{\e}{\text{e}}
\newcommand{\im}{\text{i}}
\newcommand{\be}{\begin{equation}}
\newcommand{\ee}{\end{equation}}
\newcommand{\bea}{\begin{eqnarray}}
\newcommand{\eea}{\end{eqnarray}}
\def\l{\left}
\def\r{\right}
\def\12{\frac{1}{2}}

\def\d{d}

\def\g{\gamma}

\def\l{\left}
\def\r{\right}
\def\e{\mbox{e}}

\begin{document}
\bibliographystyle{phys}
\newcommand{\onlinecite}[1]{\hspace{-1 ex} \nocite{#1}\citenum{#1}}

\markboth{J. Sirker}
{The Luttinger liquid and integrable models}

%
\catchline{}{}{}{}{}
%

\title{The Luttinger liquid and integrable models}

\author{J. Sirker}

\address{Department of Physics and Research Center OPTIMAS\\
Technical University Kaiserslautern\\
 D-67663 Kaiserslautern, Germany}



\maketitle

\begin{history}
\received{Day Month Year}
\revised{Day Month Year}
\end{history}

\begin{abstract}
  Many fundamental one-dimensional lattice models such as the
  Heisenberg or the Hubbard model are integrable. For these
  microscopic models, parameters in the Luttinger liquid theory can
  often be fixed and parameter-free results at low energies for many
  physical quantities such as dynamical correlation functions obtained
  where exact results are still out of reach. Quantum integrable
  models thus provide an important testing ground for low-energy
  Luttinger liquid physics. They are, furthermore, also very
  interesting in their own right and show, for example, peculiar
  transport and thermalization properties.  The consequences of the
  conservation laws leading to integrability for the structure of the
  low-energy effective theory have, however, not fully been explored
  yet. I will discuss the connection between integrability and
  Luttinger liquid theory here, using the anisotropic Heisenberg model
  as an example. In particular, I will review the methods which allow
  to fix free parameters in the Luttinger model with the help of the
  Bethe ansatz solution.  As applications, parameter-free results for
  the susceptibility in the presence of non-magnetic impurities, for
  spin transport, and for the spin-lattice relaxation rate are
  discussed.
\end{abstract}


\section{Introduction}
The Tomonaga-Luttinger liquid
\cite{Luttinger,MattisLieb,LutherPeschel,GiamarchiBook} is believed to
describe the low-energy properties of gapless one-dimensional
interacting electron systems irrespective of the precise nature of the
microscopic Hamiltonian. This {\it universality} can be understood in
a renormalization group sense as irrelevance of band curvature and
additional interaction terms which might arise when deriving this
low-energy effective theory from a microscopic model. Similar to the
important role Onsager's exact solution\cite{Onsager} of the
two-dimensional Ising model has played in establishing and confirming
general renormalization group theory, exactly solvable one-dimensional
quantum models have been crucial for the development of Luttinger
liquid theory.

Integrable models are, furthermore, also interesting in their own
right and a number of almost ideal realizations are known today. One
example are cuprate spin chains such as Sr$_2$CuO$_3$ whose magnetic
properties are well described by the integrable one-dimensional
Heisenberg
model.\cite{MotoyamaEisaki,TakigawaMotoyama,ThurberHunt,EggertSrCuO,SirkerLaflorencie,SirkerLaflorencie2,SirkerLaflorencieEPL,SirkerPereira,SirkerPereira2}
Furthermore, cold atomic gases represent quantum systems which are to
a high degree isolated from the surroundings and whose Hamiltonians
are easily tunable. This makes it possible to use them as quantum
simulators to study almost perfect realizations of integrable systems
such as the Lieb-Liniger model\cite{LiebLiniger,KinoshitaWenger} or
the fermionic Hubbard model.\cite{YangYang}

In Sec.\ref{QI} I will discuss quantum integrability with a particular
emphasis on Bethe ansatz integrable models and outline possible
effects on transport and the thermalization of closed quantum systems.
In the rest of the paper, I will then concentrate on the anisotropic
Heisenberg (or $XXZ$) model as one specific example for a Bethe ansatz
integrable model. In Sec.~\ref{LL} I describe how the Luttinger model,
including leading irrelevant operators, can be obtained from this
microscopic model by using bosonization techniques.  In Sec.~\ref{BA}
I then briefly outline important aspects of the Bethe ansatz solution.
In Sec.~\ref{Parameters} it is shown that a comparison of the results
of Sec.~\ref{LL} and Sec.~\ref{BA} allows to fix parameters in the
Luttinger liquid theory for the $XXZ$ model.  Applications of the
parameter-free low-energy effective theory to calculate various
properties of spin chains are considered in Sec.~\ref{Appl}. This
includes the calculation of susceptibilities in the presence of
non-magnetic impurities, and results for spin transport and NMR
relaxation rates.
The final section is devoted to a brief summary and some conclusions.

\section{Quantum integrability}
\label{QI}
A classical system with Hamilton function $\mathcal{H}$ and phase
space dimension $2N$ is integrable if it has $N$ constants of motion
$\mathcal{Q}_n$ with
\begin{equation}
\label{Classical}
\{\mathcal{H},\mathcal{Q}_n\} = 0\quad \mbox{and}\quad \{\mathcal{Q}_i,\mathcal{Q}_j\} = 0 \quad\mbox{if} \quad i\neq j \, .
\end{equation}
Here $\{.,.\}$ denotes the Poisson bracket. Quantum integrability, on
the other hand, is much harder to define precisely, see, for example,
Ref.~\citeonline{CauxMossel}. In this regard it is important to note that
every quantum system in the thermodynamic limit, irrespective of
integrability, has infinitely many conservation laws
\begin{equation}
\label{Projectors}
[H,|E_n\rangle\langle E_n|] = 0
\end{equation}
where $H$ is the Hamiltonian with eigenstates $|E_n\rangle$ and
$[.,.]$ denotes the commutator. Apart from these {\it non-local}
conservation laws a quantum system can have local conservation laws
given by
\begin{equation}
\label{local_cons}
\mathcal{Q}_n =\sum_j q_{n,j} \quad \mbox{or} \quad \mathcal{Q}_n= \int dx\, q_n(x)
\end{equation}
where $q_{n,j}$ is a density operator acting on $n$ neighboring sites
in the case of a lattice model while for a continuum model $q_n(x)$ is
a fully local density operator. A generic example for a local
conservation law is the Hamiltonian itself for models with short range
interactions. In Bethe ansatz integrable models a whole set of such
local conservation laws does exist which can be obtained from the
transfer matrix of the corresponding two-dimensional classical model
by taking successive derivatives of the transfer matrix $\tau$ with
respect to the spectral parameter $\lambda$
\begin{equation}
\label{transfer_matrix}
\mathcal{Q}_n \propto\frac{\partial^n}{\partial\lambda^n} \ln \tau(\lambda)\bigg|_{\lambda=\xi}\, .
\end{equation}
Here $\xi$ is the spectral parameter at which the transfer matrix is
evaluated. These conserved quantities are directly related to the
existence of so-called $R$-matrices which fulfill the Yang-Baxter
equations and from which the transfer matrices $\tau(\lambda)$ can be
constructed.\cite{HubbardBook}

For a low-energy effective theory describing such an integrable model,
we have to demand---at least in principle---that the low-energy Hamiltonian
$H_{\rm eff}$ also fulfills
\begin{equation}
\label{H_and_Q}
[H_{\rm eff},\mathcal{Q}_n] = 0.
\end{equation}
This corresponds to a fine-tuning of parameters in the Luttinger
model. In particular, it might mean that certain terms which are not
forbidden by general symmetry considerations have to vanish. Such a
program has not fully been explored yet; in Sec.~\ref{coup_const} we
will see, as an example, that the conserved quantity $\mathcal{Q}_3$ for
the $XXZ$ model does indeed prevent certain terms from occuring in the
low-energy theory.

\subsection{Consequences for transport}
\label{Intro_Transport}
Local conservation laws can have a dramatic effect on the transport
properties.\cite{ZotosPrelovsek} This can be easily understood as
follows. 
We can always define a local current density $j_l$ by making use of
the continuity equation
\begin{equation}
\label{Cont_eq}
\frac{\partial}{\partial t} \rho_l + j_{l} -j_{l-1} =0  
\end{equation}
where $\rho_l$ is the density at site $l$. The current itself is then
given by $\mathcal{J}=\sum_l j_l$. This current could be, for example,
an electric, spin or thermal current. Conserved quantities $Q_n$ can
now prevent a current from decaying completely leading to ballistic
transport and a {\it finite Drude weight}
\begin{equation}
\label{Drude}
D(T) \equiv \lim_{t\to\infty}\lim_{L\to\infty} \frac{1}{2LT}\langle\mathcal{J}(t)\mathcal{J}(0)\rangle \geq\lim_{L\to\infty}\frac{1}{2LT}\sum_n \frac{\langle \mathcal{J}Q_n\rangle^2}{\langle Q_n^2\rangle} \,. 
\end{equation} 
Here $Q_n$ can denote a local or non-local conserved quantity, $L$ is
the length of the system, and $T$ the temperature. The second relation
in Eq.~(\ref{Drude}) is the Mazur inequality\cite{Mazur} which becomes
an equality if all conservation laws, local and non-local, are
included.\cite{Suzuki,JungRoschlowerbound} In order to obtain a
possible non-zero Drude weight at finite temperatures within a
Luttinger model description, the relevant conservation laws have to be
taken into account explicitly. One way to achieve this is discussed in
Sec.~\ref{Transport}. Importantly, one expects that only local or
pseudo-local conservation laws $\mathcal{Q}_n$ with
$\langle\mathcal{J}\mathcal{Q}_n\rangle\neq 0$ can give rise to a
finite bound in Eq.~(\ref{Drude}) so that $D(T>0)\neq 0$ is
characteristic for an integrable model.

\subsection{Consequences for thermalization in closed systems}
Additional local conservation laws can also have a profound impact on a
possible thermalization of a closed quantum system. Imagine that we
prepare an initial state $|\Psi(0)\rangle$ and follow the unitary time
evolution of this state under an integrable Hamiltonian. One says that
a closed quantum system in the thermodynamic limit has thermalized if
for any {\it local} observable $\mathcal{O}$ the limit
\begin{equation}
\label{therm1}
\mathcal{O}_\infty = \lim_{t\to\infty} \langle \Psi(t) | \mathcal{O} | \Psi(t)\rangle
\end{equation}
is well-defined and time independent and can also be expressed as an
ensemble average
\begin{equation}
\label{therm2}
\mathcal{O}_\infty \equiv \mbox{Tr}(\mathcal{O}\rho)
\end{equation}
with an appropriately chosen density matrix $\rho$. Note that even for
a generic closed quantum system temperature $T$ is not defined by an
external bath but rather by the energy of the initial state
\begin{equation}
\label{therm3}
\langle\Psi(0)|H|\Psi(0)\rangle = \underbrace{\mbox{Tr}(H\e^{-H/T})/Z}_{\mbox{Tr}(\rho_c H)}
\end{equation}
with $T$ acting as a Lagrange multiplier and $Z$ being the partition
function. For an integrable model we have to demand that the relation
(\ref{therm3}) also holds if we replace $H\to\mathcal{Q}_n$ and
the canonical density matrix $\rho_c$ by the density matrix\cite{Deutsch}
\begin{equation}
\label{therm4}
\rho =\frac{1}{Z}\exp\bigg(-\sum_j\lambda_j \mathcal{Q}_j\bigg)
\end{equation}
which now contains a Lagrange multiplier $\lambda_j$ for {\it each} of
the locally conserved quantities. The existence of additional local
conservation laws therefore severely restricts a possible
thermalization of the system leading to additional constraints which
are incorporated by the Lagrange multipliers in Eq.~(\ref{therm4}).
Experimental indications for such constraints have been seen in
realizations of the Lieb-Liniger model in ultracold
gases.\cite{KinoshitaWenger}

\section{Low-energy description of the $XXZ$ model}
\label{LL}
In the following sections, we want to concentrate on one of the simplest integrable lattice models, the $XXZ$ model 
\begin{equation}
\label{XXZ_fermion}
H=J\sum_{j=1}^{N(N-1)}\left[-\frac{1}{2}\left(c_{j}^{\dagger}c_{j+1}^{\phantom{\dagger}}+h.c.\right)-h\left(c_{j}^{\dagger}c_{j}^{\phantom{\dagger}}-\frac{1}{2}\right)+\Delta\left(n_{j}-\frac{1}{2}\right)\left(n_{j+1}-\frac{1}{2}\right)\right].
\end{equation}
Here $c$ ($c^\dagger$) annihilates (creates) a spinless fermion, $J$
gives the energy scale, $\Delta$ characterizes the nearest neighbor
density-density interaction with the density operator $n_j=c^\dagger_j
c_j$.  $N$ is the number of sites and the boundary conditions might be
either periodic (sum runs up to $N$ with $c^{(\dagger)}_{N+1}\equiv
c^{(\dagger)}_1$) or open (sum runs only up to $N-1$). $h$ acts as a
chemical potential. With the help of the Jordan-Wigner transformation
\begin{equation}
S_{j}^{z} \rightarrow  n_{j}-\frac{1}{2},\quad
S_{j}^{+} \rightarrow  \left(-1\right)^{j}\, c_{j}^{\dagger}e^{i\pi\phi_{j}}, \quad
S_{j}^{-} \rightarrow  \left(-1\right)^{j}\, c_{j}^{\phantom{\dagger}}e^{-i\pi\phi_{j}},
\label{eq:WignerJordan}
\end{equation}
where $\phi_j=\sum_{l=1}^{j-1}n_l$ we can also express this model in
terms of spin-$1/2$ operators
\begin{equation}
\label{XXZ}
H=J\sum_{j=1}^{N(N-1)} [S^x_jS^x_{j+1}+S^y_jS^y_{j+1}+\Delta S^z_jS^z_{j+1}-hS^z_j] \, .
\end{equation}
For both kinds of boundary conditions the $XXZ$ model is integrable by
Bethe ansatz.\cite{Bethe,YangYang,Sklyanin,AlcarazBarber,tak99}
The Luttinger liquid approach is applicable in the critical regime
which is given by $-1<\Delta\leq 1$ for $h=0$.  In general, the range
of anisotropies for which the model is critical depends on the applied
magnetic field $h$. In the free fermion case, the model
(\ref{XXZ_fermion}) is easily solved by Fourier transform leading to
\begin{equation}
\label{XX}
H_0=\sum_p\epsilon_p c_p^\dagger c_p \quad \mbox{with} \quad \epsilon_p=-J(\cos p +h) \, .
\end{equation}
where we have set the lattice constant $a=1$. The allowed momenta are
given by $p=2\pi n/N$ with $n=0,\cdots,N-1$ for periodic boundary
conditions (PBCs) or $p=\pi n/(N+1)$, $n=1,\cdots, N$ for open
boundary conditions (OBCs). In Sec.~\ref{BA} we will briefly discuss
the Bethe ansatz solution of this model for OBCs.


Let us first revisit the derivation of an effective low-energy
description, the Luttinger theory, by bosonization following
Refs.~\citeonline{EggertAffleck92,GiamarchiBook,PereiraSirkerJSTAT}.
First, we replace the fermionic operators in the continuum limit by
two fields $\psi_{R,L}$ defined near the two Fermi points $\pm
k_F=\pm\arccos(-h)$:
\begin{equation}
c_{j}\rightarrow\psi\left(x\right)=e^{ik_{F}x}\psi_{R}\left(x\right)+e^{-ik_{F}x}\psi_{L}\left(x\right).\end{equation}
In a second step, we use standard Abelian bosonization to write the fermion fields as
\begin{equation}
\psi_{R,L}\left(x\right)\sim\frac{1}{\sqrt{2\pi\alpha}}\, e^{-i\sqrt{2\pi}\phi_{R,L}\left(x\right)},\label{eq:mandelstam}
\end{equation}
where $\alpha\sim k_F^{-1}$ is a short distance cutoff. Instead of
working with the left and right components $\phi_{R,L}$ we can define
a bosonic field $\tilde\phi$ and its dual field $\tilde\theta$ by
\begin{equation}
\tilde{\phi} =  \frac{\phi_{L}-\phi_{R}}{\sqrt{2}},\qquad
\tilde{\theta}  =  \frac{\phi_{L}+\phi_{R}}{\sqrt{2}},
\label{eq:thetatilde}
\end{equation}
which satisfy the standard bosonic commutation rule
$[\tilde{\phi}\left(x\right),\partial_{x^{\prime}}\tilde{\theta}\left(x^{\prime}\right)]=i\delta\left(x-x^{\prime}\right)$.

If we bosonize the kinetic energy term of Eq.~(\ref{XXZ_fermion})
keeping only the lowest order we obtain
\begin{equation}
  H_{0}^{kin}  =   iv_{F}\int_{0}^{L}\!\! dx\,\left(:\psi_{R}^{\dagger}\partial_{x}\psi_{R}^{\phantom{\dagger}}:\,-\,:\psi_{L}^{\dagger}\partial_{x}\psi_{L}^{\phantom{\dagger}}:\right) 
  =  \frac{v_{F}}{2}\int_{0}^{L}\!\! dx\,\left[\left(\partial_{x}\phi_{R}\right)^{2}+\left(\partial_{x}\phi_{L}\right)^{2}\right]
\label{eq:boskin}
\end{equation}
where $:\, :$ denotes normal ordering and $v_F=J\sin k_F$ is the Fermi
velocity. This approximation corresponds to a linearization of the
dispersion $\epsilon_p$ at the Fermi points $\pm k_F$. In this case
the bosonic model is quadratic in $\partial_x\phi_{R,L}$. Corrections
to the kinetic energy appear due to band curvature. Including these
curvature terms, we can write the expansion of the dispersion
$\epsilon_p$ near the two Fermi points as
\begin{equation}
  \epsilon_{k}^{R,L}\approx\pm v_{F}k+\frac{k^{2}}{2M}\mp\frac{\gamma k^3}{6}+\dots,
\label{definevmgamma}
\end{equation}
where $k\equiv p\mp k_{F}$ for the right or left movers, respectively,
$M=\left(J\cos k_{F}\right)^{-1}$ is the effective mass and
$\gamma=J\sin k_F$. Note that the inverse mass $M^{-1}$ vanishes in
the particle-hole symmetric case $h=0$. In this case, the curvature
correction is cubic in momentum. Bosonization of the $k^2$-term leads
to a correction cubic in $\partial_x\phi_{R,L}$, whereas the term
cubic im momentum gives a quartic correction in terms of the bosonic
fields. Cubic and quartic terms in the bosonic operators will also
arise from the interaction term in Eq.~(\ref{XXZ_fermion}). The
scaling dimension of these terms is $3$ respectively $4$ so that they
are formally irrelevant. The interaction will, however, also produce
additional marginal terms, quadratic in the bosonic fields, which
together with (\ref{eq:boskin}) lead to the exactly solvable Luttinger
model \be H_{LL}=\frac{v_{F}}{2}\int dx\left\{
  \left(1+\frac{g_{4}}{2\pi
      v_{F}}\right)\left[\left(\partial_{x}\phi_{R}\right)^{2}+\left(\partial_{x}\phi_{L}\right)^{2}\right]\right. 
\left. -\frac{g_{2}}{\pi
    v_{F}}\,\partial_{x}\phi_{L}\partial_{x}\phi_{R}\right\} .
\label{eq:quadratic}
\ee Here $g_{2}=g_{4}=2J\Delta[1-\cos(2k_{F})]=4J\Delta\sin^{2}k_{F}$
are interaction parameters. The Hamiltonian (\ref{eq:quadratic}) can
be rewritten in the form\begin{equation} 
H_{LL}=\frac{1}{2}\int
  dx\,\left[vK\left(\partial_{x}\tilde{\theta}\right)^{2}+\frac{v}{K}\left(\partial_{x}\tilde{\phi}\right)^{2}\right],
\label{eq:luttingermodel}
\end{equation}
where $v$ (the renormalized velocity) and $K$ (the Luttinger
parameter) are given by
\begin{eqnarray}
  v & = & v_{F}\sqrt{\left(1+\frac{g_{4}}{2\pi v_{F}}\right)^{2}-\left(\frac{g_{2}}{2\pi v_{F}}\right)^{2}}\approx v_{F}\left(1+\frac{2\Delta}{\pi}\sin k_{F}\right),\label{eq:velocity}\\
  K & = & \sqrt{\frac{1+\frac{g_{4}}{2\pi v_{F}}-\frac{g_{2}}{2\pi
        v_{F}}}{1+\frac{g_{4}}{2\pi v_{F}}+\frac{g_{2}}{2\pi
        v_{F}}}}\approx1-\frac{2\Delta}{\pi}\sin
  k_{F}.\label{eq:luttingerK}
\end{eqnarray} 
Expressions (\ref{eq:velocity}) and (\ref{eq:luttingerK}) are
approximations valid in the limit $|\Delta|\ll 1$. In
Sec.~\ref{Parameters} we will review how these parameters in the
Luttinger liquid Hamiltonian can be fixed exactly for arbitrary
interaction strengths $-1<\Delta\leq 1$ using the Bethe ansatz
solution.

The Luttinger parameter in the Hamiltonian (\ref{eq:luttingermodel})
can be absorbed by performing a canonical transformation that rescales
the fields in the form $\tilde{\phi}\rightarrow\sqrt{K}\phi$ and
$\tilde{\theta}\rightarrow\theta/\sqrt{K}$ leading to
\begin{equation}
H_{LL}=\frac{v}{2}\int dx\,\left[\left(\partial_{x}\theta\right)^{2}+\left(\partial_{x}\phi\right)^{2}\right].\label{eq:Hluttingerrescaled}\end{equation}
 We can also define the right and left components of these rescaled
bosonic fields by\begin{equation}
\varphi_{R,L}=\frac{\theta\mp\phi}{\sqrt{2}}.\label{eq:phi_LR}\end{equation}
These are related to $\phi_{R,L}$ by a Bogoliubov transformation.

\subsection{Irrelevant operators in the finite field case}
The leading irrelevant operators stem from the $k^2$-term in
Eq.~(\ref{definevmgamma}) and give rise to dimension three operators
$\sim (\partial_x\phi_{R,L})^3$. Similar terms will also arise by
bosonizing the interaction term. Instead of deriving these terms from
the microscopic Hamiltonian, we can introduce them phenomenologically
by considering the symmetries of the problem. In particular, the
low-energy effective theory has to be symmetric under the parity
transformation $\phi_L\to\phi_R$, $\phi_R\to\phi_L$, and $x\to -x$. We
therefore can parametrize these terms as\cite{PereiraSirkerJSTAT}
\be \delta
H=\frac{\sqrt{2\pi}}{6}\int dx\,\left\{
  \eta_{-}\left[\left(\partial_{x}\varphi_{L}\right)^{3}-\left(\partial_{x}\varphi_{R}\right)^{3}\right]\right.
\left.+\eta_{+}\left[\left(\partial_{x}\varphi_{L}\right)^{2}\partial_{x}\varphi_{R}-\left(\partial_{x}\varphi_{R}\right)^{2}\partial_{x}\varphi_{L}\right]\right\}
.
\label{eq:deltaHzetas}
\ee 
We will see in Sec.~\ref{Parameters} that we can relate the
amplitudes $\eta_\pm$ to quantities which are known from the
exact solution. The derivation of these terms starting from the microscopic Hamiltonian, on the other hand, would only allow us to obtain the coupling constants to first order in $\Delta$ with\cite{PereiraSirkerJSTAT} 
\begin{equation}
\eta_{-}  \approx  \frac{1}{M}\left(1+\frac{2\Delta}{\pi}\sin k_{F}\right), 
\quad \eta_{+}  \approx  -\frac{3\Delta}{\pi M}\sin k_{F}.
\label{eq:zetaplus}
\end{equation}
From this expansion we see that (a) both terms vanish in the limit
$h\to 0$ where $M^{-1}\to 0$, and (b) that the term mixing right and
left movers parametrized by $\eta_+$ is only present in the
interacting case, $\Delta\neq 0$.


\subsection{Irrelevant operators for zero field}
In the particle-hole symmetric case, $h=0$, the first correction to
the linear dispersion relation is cubic in momentum, see
Eq.~(\ref{definevmgamma}). Instead of bosonizing this term starting
from the microscopic Hamiltonian (\ref{XXZ_fermion}) we again
introduce the corresponding terms in the bosonic model based on
symmetry arguments. The dimension four operators allowed by symmetry
can be parametrized as
\begin{eqnarray}
\delta \mathcal{H}&=&\frac{\pi\zeta_-}{12}\left[:\left(\partial_x\varphi_R\right)^2:\,:\left(\partial_x\varphi_R\right)^2:+:\left(\partial_x\varphi_L\right)^2:\,:\left(\partial_x\varphi_L\right)^2:\right]+\frac{\pi\zeta_+}{2}\,\left[:\left(\partial_x\varphi_R\right)^2:\right. \nonumber\\
& & \left. :\left(\partial_x\varphi_L\right)^2:\right]
+\pi\zeta_3\left[:(\partial_x\varphi_R)^3:\,:\partial_x\varphi_L: +:(\partial_x\varphi_L)^3:\,:\partial_x\varphi_R:\right]\, . 
\label{deltaH_h=0}
\end{eqnarray}
The explicit bosonization of the corresponding band curvature and
interaction terms yields the coupling constants again only to lowest
lowest order
\begin{equation}
  \zeta_-\approx -J\left(1+\frac{\Delta}{\pi}\right),\qquad
  \zeta_+\approx-\frac{\Delta J}{\pi},\qquad
  \zeta_3=0.
\end{equation}
In addition, the Umklapp scattering term $\delta\mathcal{H}_U\sim
\e^{4ik_Fx}\Psi_R^\dagger(x)\Psi_L(x)\Psi_R^\dagger(x+1)\Psi_L(x+1)+h.c.$
is commensurate in this case, $4k_F=2\pi$, and therefore has to be kept
in the low-energy effective theory. Bosonizing this term leads to
\begin{equation}
\label{H_U}
\delta\mathcal{H}_U =\lambda \cos(4\sqrt{\pi K}\phi) \, ,
\end{equation}
and to lowest order in $\Delta$ we have $\lambda=J\Delta/(2\pi^2)$.
For OBC, there is also an irrelevant boundary operator allowed 
\begin{equation}
\label{H_B}
\delta\mathcal{H}_B \sim (\delta(x)+\delta(L))(\partial_x\phi)^2 \, .
\end{equation}

Finally, we want to consider the Luttinger model
(\ref{eq:Hluttingerrescaled}) with an additional small magnetic field
$\delta h$ added, ignoring the irrelevant terms
\begin{equation}
\label{HLL_with_h}
H=\frac{v}{2}\int_0^L \left[(\partial_x\phi)^2+(\partial_x\theta)^2-\frac{2}{v}\sqrt{\frac{K}{\pi}}\delta h\partial_x\phi\right] \, .
\end{equation}
By performing a shift in the boson field 
\be
\label{Shift}
\phi\to\phi +\sqrt{\frac{K}{\pi}}\frac{x}{v} \delta h \ee we return to
the quadratic Hamiltonian (\ref{eq:Hluttingerrescaled}) with an
additional constant shift $-LK(\delta h)^2/(2\pi v)$.  The bulk
susceptibility per site is therefore given by
\begin{equation}
\label{chi}
\chi_{\rm bulk} = K/\pi v \, .
\end{equation}
This result does not only hold for $h=0$ but also for any finite field
$h_0$ at which we want to calculate $\chi$ with $K=K(\Delta, h_0)$ and
$v=v(\Delta,h_0)$.

\section{The Bethe ansatz solution}
\label{BA}
To exactly solve the interacting system for PBC or OBC, one can use
the coordinate Bethe
ansatz.\cite{Bethe,YangYang,tak99,AlcarazBarber,Sklyanin} Here we want
to review very briefly some of the essential results needed to fix the
parameters in the Luttinger model and refer the reader to
Refs.~\citeonline{tak99,Lukyanov,BortzSirker,SirkerBortzJSTAT} for a
more detailed discussion. The coordinate Bethe ansatz starts from the
fully polarized state ('the vacuum') and one derives coupled
eigenvalue equations $H|M\rangle =E|M\rangle$ for states $|M\rangle$
with $M$ spins flipped. The eigenenergies can then be written as
\begin{equation}
\label{XXZ_BA}
E=J\sum_{j=1}^M \cos k_j +J\Delta\left(\frac{N-1}{4}-M\right) \, .
\end{equation}
The structure is similar to the non-interacting case, however, the
momenta $k_j$ are shifted from their positions for $\Delta=0$. They
can be determined from a set of coupled nonlinear equations. In the
thermodynamic limit, a single integral equation for the density of
roots $\rho(x)$ is obtained which parametrizes the allowed momenta
\be
\vartheta(x,\g)+\frac{1}{2N}\l[\vartheta(x,\g)+\vartheta(x,\pi-\g)+\vartheta(x,2\g)\r]
=\rho(x)+\int_{-B}^B\vartheta(x-y,2\g)\rho(y)\,\d y
\label{XXZ_BA2},
\ee where 
\be \vartheta(x,\g)=\frac{1}{\pi}\frac{ \sin\g}{\cosh
  2x-\cos\g}  
\label{XXZ_BA3}
\ee and we have set $\Delta=\cos\gamma$. Eq.~(\ref{XXZ_BA2}) is the
integral equation for the $XXZ$ chain with OBC in the thermodynamic
limit, including the boundary correction $\sim\mathcal{O}(1/N)$.
Omitting the $1/N$ correction this is the standard integral equation
for PBC.\cite{tak99} The integral equation contains an unknown
parameter $B$ and an unknown function $\rho(x)$. It can be solved
analytically by Fourier transform for $B=\infty$ and one finds
(ignoring the $1/N$ correction) \be \rho(x)=\frac{1}{2\gamma\cosh\pi
  x/\gamma} \, .
\label{XXZ_BA4}
\ee

The magnetization per site $m$ and the ground state energy per site
$e$ for general $B$ are given by \bea
m&=& 1/2-\int_{-B}^B\rho(x)\d x+1/(2N)\nonumber \\
e&=&-h s^z-\frac{J\sin\g}{2}\int_{-B}^B \vartheta(x,\g)\rho(x)\d
x+\frac{J}{4}\l(\cos\g+\frac{2-\cos \g}{N}\r)
\label{sz2}.  
\eea
Inserting (\ref{XXZ_BA4}) into Eq.~(\ref{sz2}) one finds that
$B=\infty$ corresponds to $m=0$, i.e., to the case of zero magnetic
field $h=0$. In general, $B=B(h)$ and the dependence on magnetic field
has to be determined numerically. This can be achieved by using the
stationarity condition
\be
\frac{\partial}{\partial B}[e(h)-e(h=0)]=0
\label{var}.  
\ee

\section{Fixing parameters of the Luttinger model using integrability}
\label{Parameters}
One of the main motivations to apply Luttinger liquid theory to
integrable models is that parameters in the Luttinger liquid theory
such as velocities, Luttinger parameters, coupling constants of
irrelevant operators and prefactors of correlation functions which
usually are non-universal and therefore unknown, can often be
determined in the case of an integrable
model.\cite{Lukyanov,LukyanovTerras,KitanineMaillet,KitanineMaillet2,KitanineKozlowski}
This makes it possible to obtain parameter-free results at low
energies. The general idea is to calculate static observables at zero
or finite temperatures exactly using the Bethe ansatz and to compare
with results obtained within Luttinger theory. In the following we
briefly review this method to obtain the velocity and Luttinger liquid
parameter, Sec.~\ref{v_and_L}, and the coupling constants for band
curvature and Umklapp terms for the $XXZ$ model,
Sec.~\ref{coup_const}.

\subsection{Velocity and Luttinger liquid parameter}
\label{v_and_L}
To obtain the velocity of elementary excitations, we have to consider
the change in energy when replacing the ground state distribution of
roots by a distribution which contains an excitation near the Fermi
points. In the case of finite magnetic field, the obtained Bethe
ansatz equations can only be solved numerically. Here we want to
restrict ourselves to the zero field case. Replacing the ground state
distribution (\ref{XXZ_BA4}) in the expression for the energy
(\ref{sz2}) by the distribution containing an excitation gives the
energy in terms of the momentum change.\cite{tak99} This allows to
read off the spin velocity \be
\label{VL1}
v=\frac{\partial E}{\partial k}\bigg|_{k=k_F}
=\frac{J\pi}{2}\frac{\sin\gamma}{\gamma}=\frac{J\pi}{2}\frac{\sqrt{1-\Delta^2}}{\arccos\Delta}
\,.  \ee The spin velocity therefore increases from $v_F=J$ (remember
that we have set $a=1$) at the free fermion point to $v=J\pi/2$ at the
isotropic antiferromagnetic Heisenberg point. Conversely, the velocity
vanishes, as expected, for $\Delta\to -1$ corresponding to the
isotropic ferromagnet.

\begin{figure}
\begin{center}
\includegraphics*[width=0.75\columnwidth]{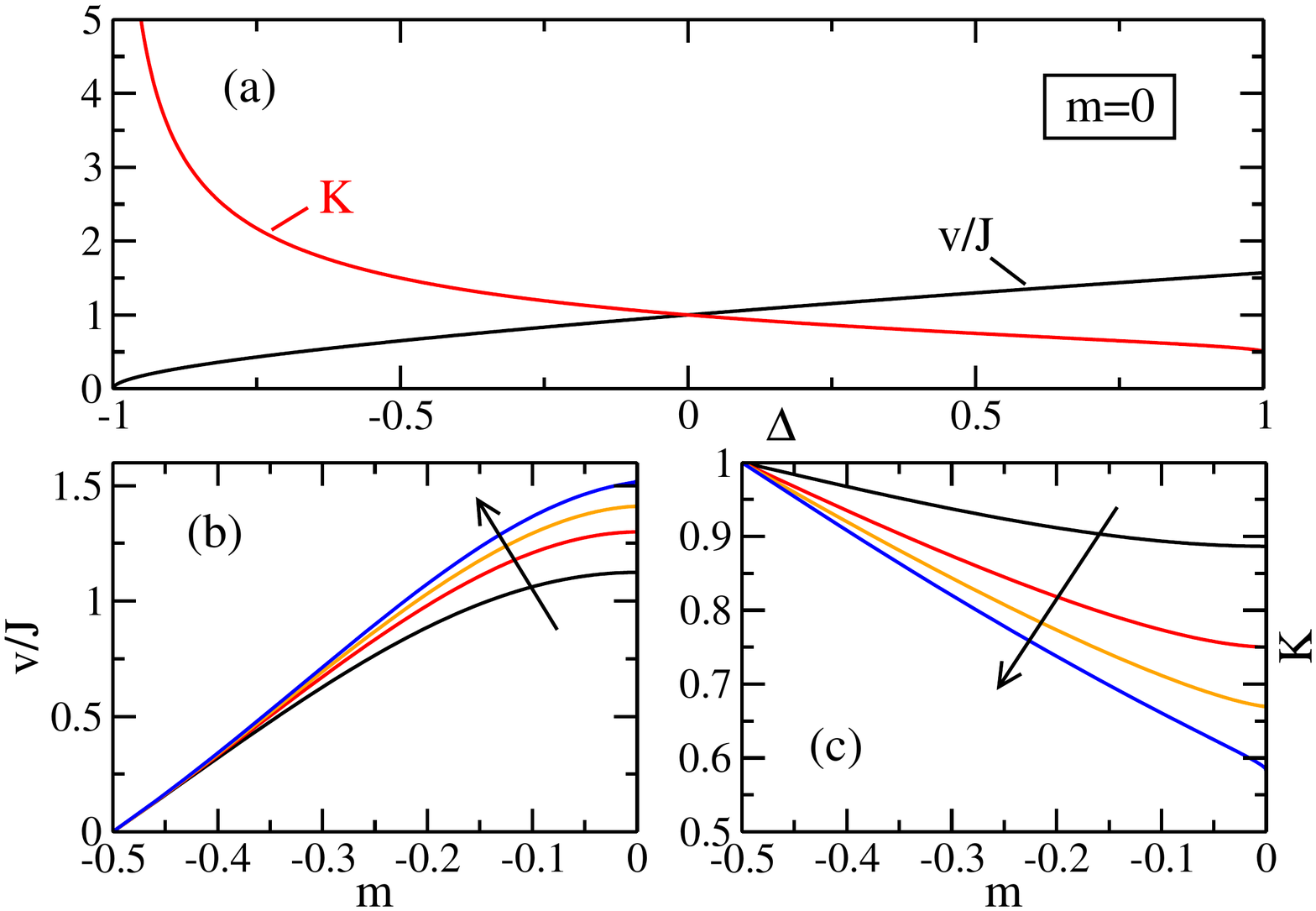}
\end{center}
\caption{(a) Velocity $v/J$ and Luttinger parameter $K$ for
  magnetization $m=0$ as a function of $\Delta$. Velocity (b) and
  Luttinger parameter (c) as a function of magnetization $m$ for
  $\Delta=0.2,0.5,0.7,0.9$ (in arrow direction).}
\label{v_and_K}
\end{figure}

To determine the Luttinger parameter $K$ it is easiest to calculate
the bulk susceptibility $\chi_{\rm bulk}(h)=\partial m/\partial h$
using Eq.~(\ref{sz2}). To do so, $B(h)$ is required. For finite
magnetic fields this again requires a numerical solution. For
infinitesimal fields, on the other hand, $B(h)$ can be determined
analytically \cite{tak99,BortzSirker,SirkerBortzJSTAT} and 
 \be \chi_{\mbox{\footnotesize bulk}}=\frac{1}{J}\frac{\g}{(\pi-\g)\pi
  \sin\g} =\frac{1}{2v(\pi-\g)} 
\label{chib}
\ee
is obtained. Comparing with Eq.~(\ref{chi}) we find
\be
\label{K}
K=\frac{\pi}{2(\pi-\gamma)}=\frac{\pi}{2(\pi-\arccos\Delta)} \, .  \ee
Therefore $K=1$ for the free fermion model and $K=1/2$ at the
isotropic antiferromagnetic Heisenberg point. Eq.~(\ref{VL1}) and
Eq.~(\ref{K}) agree to first order with the expressions
(\ref{eq:luttingerK}). The velocity and Luttinger parameter, both for zero and finite fields, as obtained from the Bethe ansatz solution, are shown in Fig.~\ref{v_and_K}.

\subsection{Coupling constants of irrelevant operators}
\label{coup_const}
Next, we want to review how the coupling constants $\zeta_{\pm,3}$,
Eq.~(\ref{deltaH_h=0}), and $\lambda$, Eq.~(\ref{H_U}), can be fixed
in the zero field case, and the coupling constants $\eta_{\pm}$,
Eq.~(\ref{eq:deltaHzetas}), for finite field. The zero field case has
been first considered by Lukyanov\cite{Lukyanov} and analytical
formulas for the coupling constants have been obtained. The finite
field case has been treated in Refs.~\citeonline{PereiraSirker,PereiraSirkerJSTAT}
leading to formulas which require a numerical solution of the Bethe
ansatz equations.
\subsubsection{The zero field case}
The simplest way to determine the Umklapp scattering amplitude
$\lambda$ is to consider an open chain with a small magnetic field
added.\cite{SirkerBortzJSTAT} In the low-energy description this means
that we have to consider (\ref{HLL_with_h}) with the Umklapp term
(\ref{H_U}) added.  We can then again perform the shift (\ref{Shift}).
This brings us back to the standard Luttinger liquid Hamiltonian
(\ref{eq:Hluttingerrescaled}) and the magnetic field now appears in
the Umklapp term (\ref{H_U}). In first order perturbation theory in
Umklapp scattering we then find the following boundary correction to
the ground state energy\cite{FujimotoEggert}
\begin{equation}
\label{B71}
E_U^{(1)} = \lambda \int_0^\infty dx \l\langle\cos\l(4\sqrt{\pi K}\phi+\frac{4Khx}{v}\r)\r\rangle_0 \, . 
\end{equation}
Here $\langle\cdots\rangle_0$ denotes the correlation function
calculated for the free boson model. For PBC this correlation function
would vanish, however, for OBC we obtain
\begin{equation}
\label{B72}
E_U^{(1)} = \lambda\int_0^\infty dx
\frac{\cos\l(\frac{4Khx}{v}\r)}{(2x)^{4K}} \; .
\end{equation} 
Note that this is a $1/N$ correction to the ground state energy per
site $e=E/N$. By partial integration we can split of the convergent
part and find
\begin{equation}
\label{B74}
E_U^{(1,conv)} = -\lambda
(2K)^{4K}\Gamma(-4K)\sin(2K\pi)\l(\frac{h}{v}\r)^{4K-1} \;.
\end{equation}
At the same time, we can apply the Bethe ansatz to analytically
calculate the so-called boundary susceptibility $\chi_B$ given by
$\chi=\chi_{\rm bulk}+\chi_B/N+\mathcal{O}(N^2)$ to leading orders in
$h$.\cite{SirkerBortzJSTAT,BortzSirker} The amplitude $\lambda$ of the
Umklapp term can now be found by comparing the exact result for
$\chi_B$ with Eq.~(\ref{B74}). This leads to
\begin{equation}
\label{B108}
\lambda
=\frac{2K\Gamma(2K)\sin \pi/2K}{\pi\Gamma(2-2K)}\l[\frac{\Gamma\l(1+\frac{1}{4K-2}\r)}{2\sqrt{\pi}\Gamma\l(1+\frac{K}{2K-1}\r)}\r]^{4K-2}\, .
\end{equation} 
In Ref.~\citeonline{Lukyanov} this result has been obtained first by
calculating the bulk correction to the ground state energy. Note,
however, that this requires second order perturbation theory in the
Umklapp scattering. Particular care has to be taken when considering
the isotropic antiferromagnet, $\Delta=1$. In this case, Umklapp
scattering becomes marginally irrelevant and $\lambda$ has to be
replaced by a running coupling constant which depends on the length
scale the system is considered at. In general, both the length of the
system and temperature will be of importance and the running coupling
constant $g(L,v/T)$ can be introduced by the replacements $K\to
(1+g)/2$ and $\lambda\to -g/4$. An explicit solution of the
renormalization group equations for $g$ is only possible if one of
those two length scales dominates. In the thermodynamic limit, for
example, this scale will be set by temperature alone and one
finds\cite{Lukyanov}
\begin{equation}
\label{running_r}
1/g+\ln(g)/2=\ln(T_0/T)
\end{equation}
with $T_0=\sqrt{\pi/2}\e^{1/4+\tilde\gamma}$ where $\tilde\gamma$ is
the Euler constant. The scale $T_0$ has again been fixed by comparing
with the Bethe ansatz result for the bulk susceptibility $\chi_{\rm
  bulk}$ in the isotropic case.

Here integrability has been used to fix a coupling constant. The
conservation laws underlying integrability discussed in Sec.~\ref{QI}
can, however, have an even more profound
effect.\cite{SirkerPereira,SirkerPereira2} For the $XXZ$ model the
first of the non-trivial conserved quantities is the energy
current $J^E=\mathcal{Q}_3$ given by \bea
J^E&=&J^2\sum_j\left[S_{j-1}^yS_j^zS_{j+1}^x-S_{j-1}^xS_j^zS_{j+1}^y+\Delta (S_{j-1}^xS_j^yS_{j+1}^z-S_{j-1}^zS_j^yS_{j+1}^x)\right.\nonumber\\
& &\left.+\Delta
  (S_{j-1}^zS_j^xS_{j+1}^y-S_{j-1}^yS_j^xS_{j+1}^z)\right].\label{JElattice}
\eea The latter is defined by the continuity equation of the energy
density at zero field \be j^E_{j+1}-j^E_j=- \partial_t \mathcal{H}_j =
i [\mathcal{H}_j,H],
\label{cont_eq}
\ee where $H=\sum_j \mathcal{H}_j$ is the Hamiltonian (\ref{XXZ}) with
$h=0$, PBC, and $J^E=\sum_j j^E_j$. The energy current operator for
the Luttinger model can be obtained from (\ref{cont_eq}) by taking the
continuum limit. This leads to \be
\label{JE_cont}
J^E_0=\int dx \, j^E_0(x)=
\frac{v^2}{2}\int dx \left[\left(\partial_x
    \varphi_R\right)^2-\left(\partial_x \varphi_L\right)^2\right]
=-v^2\int dx\, \partial_x\phi
\partial_x\theta.  \ee This operator is conserved, i.e.,
$[J_0^E,H_{LL}]=0$. The irrelevant operators (\ref{deltaH_h=0}) lead
to a correction of the energy current which can again be calculated
using the continuity equation (\ref{cont_eq}). To first order one
finds \be \delta J^E=\pi v\int dx \left\{\frac{\zeta_-}{3}
  \left[\left(\partial_x \varphi_R\right)^4-\left(\partial_x
      \varphi_L\right)^4\right]\right. 
\left.  +2\zeta_3\left[\left(\partial_x \varphi_R\right)^3 \partial_x
    \varphi_L-\left(\partial_x \varphi_L\right)^3 \partial_x
    \varphi_R\right]\right\}.
\label{deltaJE}
\ee For $J^E=J_0^E+\delta J^E$ to be conserved as required by
integrability, we have to require that $[J^E,H]= [J_0^E+\delta
J^E,H_{LL}+\delta H]=0$ up to the considered order. Since
$[J^0_E,H]=[J_0^E,H_{LL}+\delta H]=0$ this implies that $[\delta
J^E,H_{LL}]=0$. The $\zeta_-$-term in Eq.~(\ref{deltaJE}) does not mix
right and left movers and therefore obviously commutes with $H_{LL}$.
The $\zeta_3$-term, on the other hand, does mix the two modes and
therefore does not commute with $H_{LL}$. Integrability therefore
implies that $\zeta_3=0$. We see that apart from determining the
precise values of coupling constants in the low-energy effective
theory, there is a more fundamental consequence: Integrability
corresponds to a fine tuning of the coupling constants such that the
local conservation laws are fulfilled. In particular, terms which are
in general allowed by symmetry might be absent.

\begin{figure}
\begin{center}
\includegraphics*[width=0.75\columnwidth]{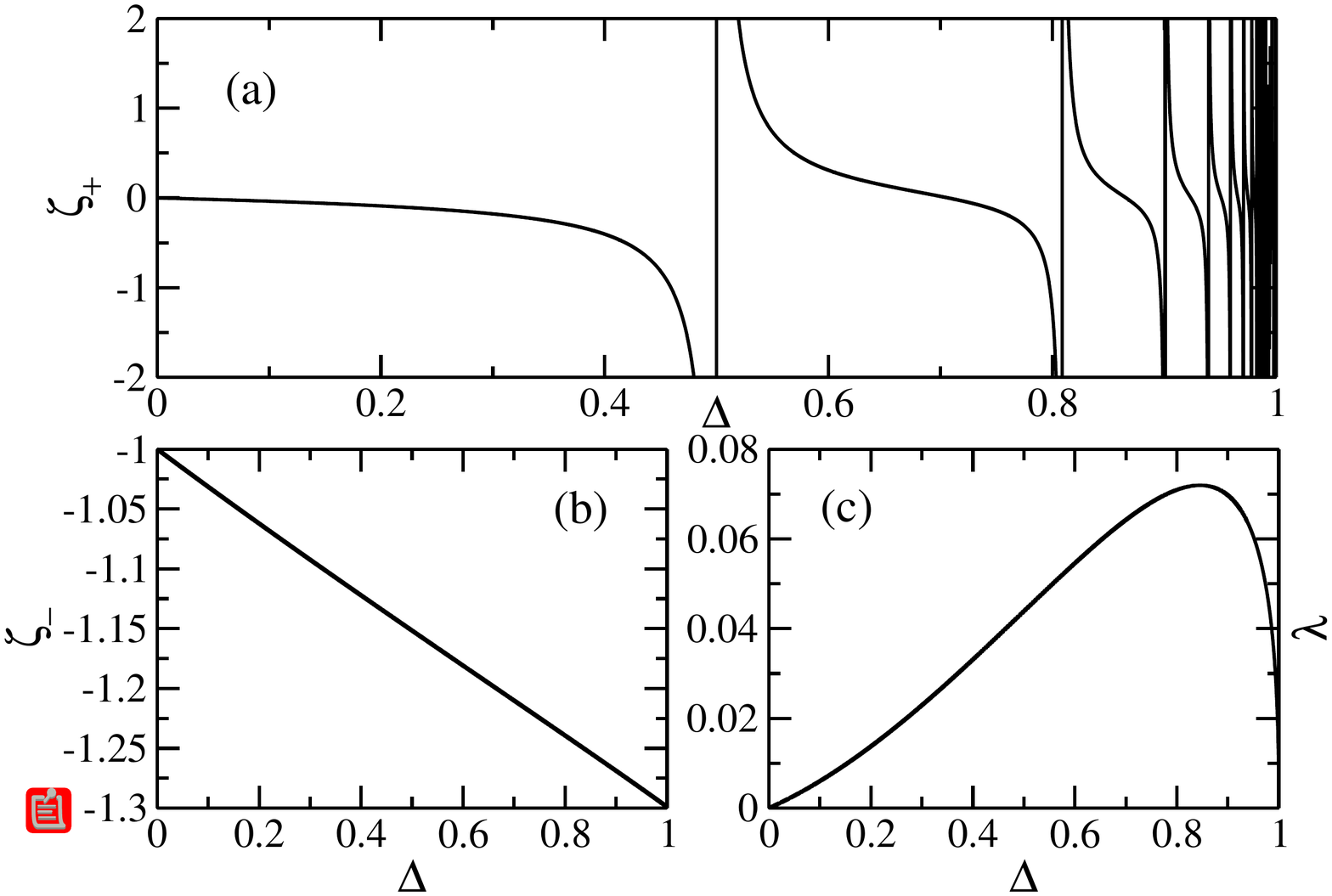}
\end{center}
\caption{(a) $\zeta_+$ as function of anisotropy $\Delta$. The
  amplitude diverges whenever the argument of the $\tan$-function in
  Eq.~(\ref{zeta+h=0}) is $\pm\pi/2 (\mbox{mod}\, 2\pi)$. (b) $\zeta_-$,
  and (c) $\lambda$, Eq.~(\ref{B108}) as a function of $\Delta$.}
\label{lambda_zetapm}
\end{figure}

We are left with only two amplitudes, $\zeta_\pm$, for the dimension
four operators. Let us briefly review how they can be fixed as well.
Using Eq.~(\ref{eq:phi_LR}) we can express both terms by the boson
field $\phi$ and the dual field $\theta$. Now performing again the
shift (\ref{Shift}) for a small applied magnetic field $\delta h$ we
find a first order correction to the ground state energy per site \be
e^{(1)}_{\zeta_\pm}=(\zeta_-+3\zeta_+)\frac{K^2(\delta h)^4}{24\pi v^4} \, .
\label{quartic_corr}
\ee The $(\delta h)^4$-term in the ground state energy can also be
calculated analytically by Bethe ansatz.\cite{Lukyanov,BortzSirker,SirkerBortzJSTAT}
One finds that the result consists of two distinct, additive,
contributions. One of those vanishes at the free fermion point and is
therefore associated with the $\zeta_+$-term in the low-energy
effective theory which mixes right and left movers. The other term
then determines $\zeta_-$ leading to\cite{Lukyanov}
\begin{equation}
\zeta_-=-\frac{v}{4\pi K}\frac{\Gamma\left(\frac{6K}{4K-2}\right)\Gamma^3\left(\frac{1}{4K-2}\right)}{\Gamma\left(\frac{3}{4K-2}\right)\Gamma^3\left(\frac{K}{2K-1}\right)},\qquad
\zeta_+=-\frac{v}{2\pi}\tan\left(\frac{\pi K}{2K-1}\right).
\label{zeta+h=0}
\end{equation}
The dependence on anisotropy of all three coupling constants
$\lambda,\zeta_\pm$ is shown in Fig.~\ref{lambda_zetapm}. Note that
the amplitude $\zeta_+$ diverges for $K=(2n+1)/4n$ with
$n\in\mathbb{N}$. Corrections to observables calculated in
perturbation theory in the irrelevant operators are, however, usually
finite. What happens is that at these special points the scaling
dimensions of different irrelevant operators coincide and the two
diverging amplitudes 'conspire' to produce a finite result. This point
has been discussed in detail in Ref.~\citeonline{SirkerBortzJSTAT}
using the susceptibility as an example.

\subsubsection{The finite field case}
For a finite magnetic field ($B<\infty$), the Bethe ansatz integral
equation cannot be solved analytically. However the amplitudes
$\eta_\pm$ can be related to changes in the Luttinger parameter and
velocity when changing the
field.\cite{PereiraSirker,PereiraSirkerJSTAT} This allows for an
accurate numerical determination of these parameters.

The basic idea is again quite simple. We consider the Luttinger liquid
Hamiltonian at some finite magnetic field $h_0$. This means that our
left and right modes live near Fermi points $\pm k_F\neq \pm \pi/2$.
Now we apply an additional small magnetic field $\delta h$ which we
can take care of by the boson shift (\ref{Shift}). If we now calculate
the free energy we obtain
\be
\label{free_E}
f=-\frac{\pi T^2}{6v(h)}-\underbrace{\frac{K(h)}{2\pi v(h)}}_{=\chi(h)/2}(J\delta h)^2 \ee with
$h=h_0+\delta h$. The interaction parameters $K(h)$ and $v(h)$ can be
determined numerically as described in Sec.~\ref{BA}. Now we can
expand (\ref{free_E}) in $\delta h$ and obtain to lowest order
\be
\delta f =\frac{\pi T^2}{6v^2(h_0)}\frac{\partial v}{\partial h}\bigg|_{h=h_0}\delta h ;
\quad
\delta\chi=\frac{K(h_0)}{\pi v(h_0)}\left[\frac{1}{K}\frac{\partial K}{\partial h}\bigg|_{h=h_0}-\frac{1}{v}\frac{\partial v}{\partial h}\bigg|_{h=h_0}\right]\delta h
\label{freeEcorr}
\ee These corrections have to stem from the dimension three operators
(\ref{eq:deltaHzetas}). The second approach therefore is to keep
$K=K(h=h_0)$, $v=v(h=h_0)$ fixed and to perform the shift
(\ref{Shift}) also in (\ref{eq:deltaHzetas}).  Calculating again the
free energy by standard techniques we now find \be \delta
f=\left(3\eta_- -\eta_+\right)\frac{\pi\,\sqrt{K}J\delta h\,
  T^{2}}{18v^3(h_0)}+(\eta_- -\eta_+)\,\frac{K^{3/2}\left(J\delta
    h\right)^{3}}{6\pi v^{3}(h_0)}; \quad
\delta\chi=(\eta_+-\eta_-)\frac{K^{3/2}J}{\pi v^3(h_0)}\delta h
\label{eq:deltaf2nd}
\ee A comparison of (\ref{freeEcorr}) and (\ref{eq:deltaf2nd}) yields
two equations from which one
obtains\cite{PereiraSirker,PereiraSirkerJSTAT}
\begin{equation}
\eta_{-} =  \frac{v}{K^{1/2}}\frac{\partial v}{\partial h}+\frac{v^{2}}{2K^{3/2}}\frac{\partial K}{\partial h},\qquad 
\eta_{+} =  \frac{3v^{2}}{2K^{3/2}}\frac{\partial K}{\partial h}.
\label{eq:identity2}
\end{equation}
From this result a few general conclusions can be drawn. For a free
model, $\eta_+$ should vanish because a mixing of right and left
movers is then impossible. This is indeed the case since $K$ does not
change when applying a field in this case.  Furthermore, $\eta_+$ will
also be absent for models such as the Calogero-Sutherland model where
$K$ remains independent of the applied field even in the interacting
case. The $\eta_-$-term, on the other hand, is already present in a
non-interacting system due to band curvature with $\eta_-=M^{-1}$ for
$\Delta=0$. The result (\ref{eq:identity2}) of course also agrees with
the expansion for small $\Delta$, Eq.~(\ref{eq:zetaplus}).

\begin{figure}
\begin{center}
\includegraphics*[width=0.75\columnwidth]{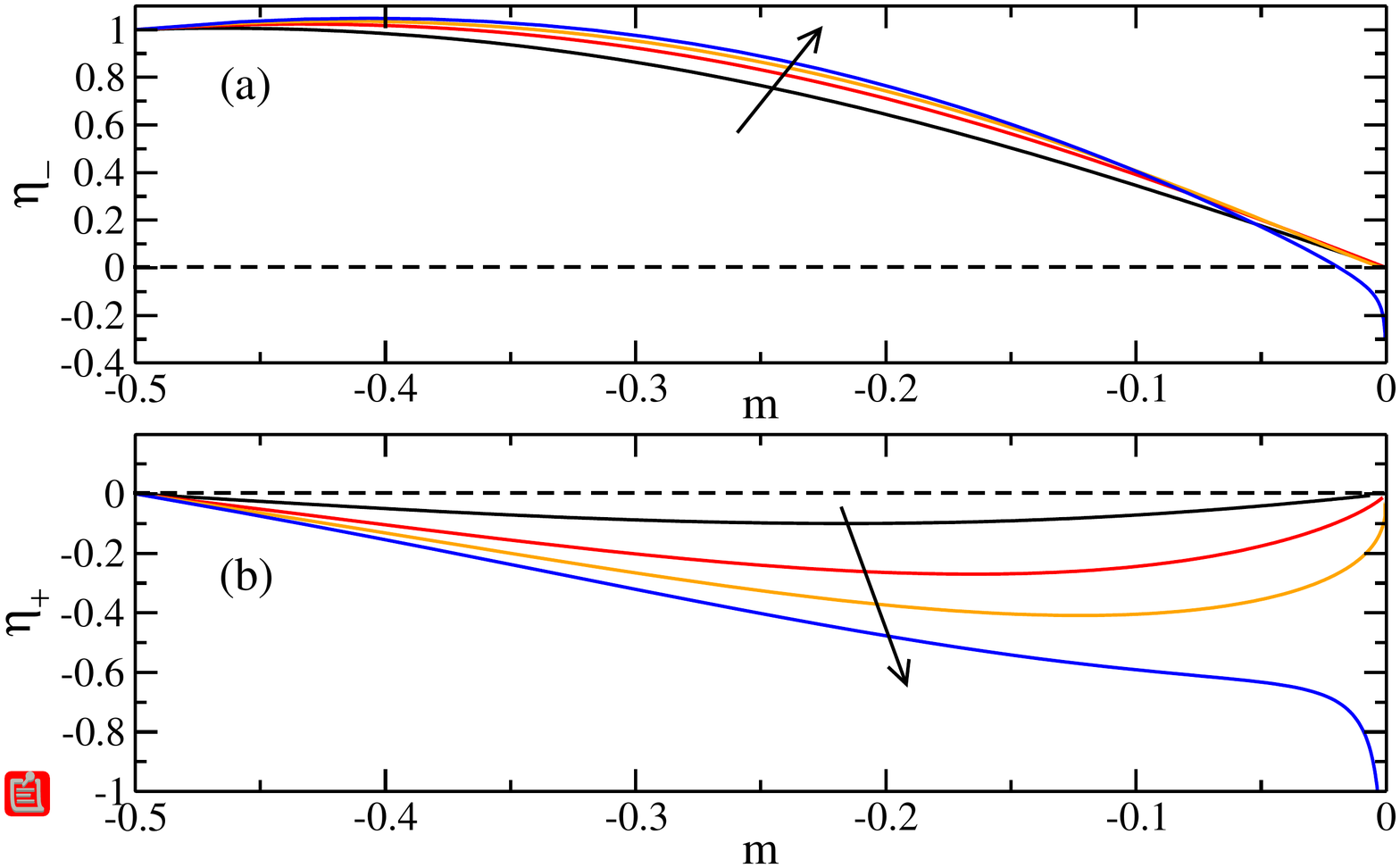}
\end{center}
\caption{(a) $\eta_-$ as a function of magnetization $m$ for
  $\Delta=0.2,0.5,0.7,0.9$ (in arrow direction). (b) Same for
  $\eta_+$. Note that $\eta_\pm\to -\infty$ for $m\to 0$ in the case
  $\Delta=0.9$ leading to a sign change of $\eta_-$.}
\label{etapm}
\end{figure}

The parameters $\eta_\pm$ are shown in Fig.~\ref{etapm} for different anisotropies $\Delta$ as a function of magnetization $m$.
While $\eta_\pm\to 0$ for $m\to 0$ if $K>5/8$ as expected from the
weak coupling expansion (\ref{eq:zetaplus}) both parameters diverge,
on the other hand, for $K<5/8$ ($\Delta>\cos(\pi/5)\approx 0.81$).
This behavior is discussed in more detail in
Ref.~\citeonline{SirkerPereira2} and we remind the reader that
divergencies also occur in the amplitude $\zeta_+$ for $h=0$.

\section{Applications}
\label{Appl}
We now want to consider a few examples where the low-energy effective
theory has been used to obtain parameter-free results for several
important observables.
\subsection{Impurities, Friedel oscillations and nuclear magnetic resonance}
\label{NMR}
One of the best known realizations of the spin-$1/2$ antiferromagnetic
Heisenberg chain is the cuprate Sr$_2$CuO$_3$.\cite{MotoyamaEisaki} In
this system excess oxygen dopes holes into the chain which seem to be
basically immobile.\cite{TakigawaMotoyama,BoucherTakigawa}
Effectively, this leads to randomly distributed non-magnetic
impurities which cut the spin chain into finite segments.  The
magnetic properties are therefore determined by an ensemble of finite
chains of random length $N$ with OBC.

In a chain with OBC, translational invariance is broken leading to a
position dependent local susceptibility
 \begin{equation}
\label{eq2} 
\chi_j=\frac{\partial}{\partial h} \langle S^z_j\rangle_{h=0} = \frac{1}{T} \langle
S^z_j S^z_{\rm tot}\rangle_{h=0}
\end{equation}
where $T$ is the temperature and $S^z_{\rm tot} = \sum_j S^z_j$. In
order to calculate $\chi_j$ in the low-energy limit, we can
express the spin operator in terms of the bosonic field
\begin{equation}
\label{eq3}
S^z_j \approx \sqrt{\frac{K}{\pi}}\partial_x\Phi + c (-1)^j \cos\sqrt{4\pi
    K}\Phi \; .
\end{equation}
Here the prefactor of the uniform part is fixed by the condition
$\sum_j S^z_j =S^z_{\rm tot}$.\cite{EggertAffleck92,GiamarchiBook} The
amplitude of the alternating part, on the other hand, can be fixed
with the help of the Bethe ansatz solution. The techniques required
are, however, much more involved than the ones reviewed in the
previous section. In particular, one finds that $c=\sqrt{A_z/2}$ with
$A_z$ as given in Eq.~(4.3) of Ref.~\citeonline{LukyanovTerras}.

Using Eq.~(\ref{eq3}) we can write $\chi_j =\chi^{\rm uni} +
(-1)^j\chi_j^{\rm st}$. The uniform part $\chi^{\rm uni}$ for the
Luttinger model is given by
\begin{equation}
\label{scalingpart}
\chi^{\rm uni} = -\frac{\partial^2 f}{\partial h^2}\bigg|_{h=0} 
=\frac{1}{LT} \frac{\sum_{S_z} S_z^2 \exp\l[-\frac{\pi
    v}{2KLT}S_z^2\r]}{\sum_{S_z} \exp\l[-\frac{\pi v}{2KLT}S_z^2\r]} \, .
\end{equation}
For $LT/v\to 0$ and $L$
even 
$\chi_s \sim \frac{2}{LT}\exp\l[-\frac{\pi v}{2KLT}\r]$ whereas for
$L$ odd $\chi_s \sim (4LT)^{-1}$. For $LT/v\to \infty$ the
thermodynamic limit result (\ref{chi}) is recovered. Note that this
zeroth order result is position independent and shows scaling with
$LT$. Corrections to scaling occur due to the irrelevant bulk and
boundary operators. For $0\leq\Delta\leq 1$ the leading bulk
irrelevant operator is due to Umklapp scattering (\ref{H_U}). This
leads to a first order correction in the free energy
\begin{equation}
\label{free_E_corr}
\delta f_1 =\frac{\lambda}{L}\int_0^L dx \langle \cos(4\sqrt{\pi K}\phi)\rangle_{S_z}\exp(-8\pi K \langle\phi\phi\rangle_{\rm osc})
\end{equation}
where we have used the mode expansion for OBC
\begin{eqnarray}
\label{ModeExp}
\phi(x,t) &=& \sqrt{\frac{\pi}{16K}} +\sqrt{\frac{\pi}{K}} S^z_{\rm tot} \frac{x}{L} + \sum_{n=1}^\infty
\frac{\sin\l(\pi nx/L\r)}{\sqrt{\pi n}}\l(\e^{-i\pi n \frac{vt}{L}}a_n +
\e^{i\pi n\frac{vt}{L}}a_n^\dagger\r) 
\end{eqnarray}
to split the expectation value of the Umklapp operator into an $S_z$
(zero mode) and an oscillator part. Furthermore, we have used the
cumulant theorem for the oscillator part. It is now straightforward,
although a bit tedious, to evaluate the two parts of
(\ref{free_E_corr}). From this the correction to the uniform part of
the susceptibility in first order in Umklapp scattering can readily be
obtained. 
\cite{SirkerLaflorencie2}



The boundary operator (\ref{H_B}) yields a further correction\cite{SirkerLaflorencie2}
\begin{equation}
\label{NonU}
\delta \chi^{\rm uni}_2 = \frac{\pi vb}{2KT^2L^3} \left[\frac{\sum_{S_z} S_z^4  \e^{-\frac{\pi vS_z^2}{2KLT}}}{\sum_{S_z}
  \e^{-\frac{\pi vS_z^2}{2KLT}}} - \frac{\l(\sum_{S_z} S_z^2  \e^{-\frac{\pi vS_z^2}{2KLT}}\r)^2}{\l(\sum_{S_z}
  \e^{-\frac{\pi vS_z^2}{2KLT}}\r)^2}\right] \, .
\end{equation}
In the thermodynamic limit, Eq.~(\ref{NonU}) reduces to $\delta
\chi^{\rm uni}_2 \to Kb/(\pi v L)$. The field theory result in this
limit can be compared with the calculation of the boundary
susceptibility based on the Bethe ansatz
\cite{BortzSirker,SirkerBortzJSTAT} and the proportionality constant
$b$ can be fixed
\begin{equation}
\label{NonU3}
b=2^{-1/2}\sin\l[\pi K/(4K-2)\r]/\cos\l[\pi/(8K-4)\r] \; .
\end{equation}
To first order in Umklapp scattering and in the dimension three
boundary operator a parameter-free result for $\chi^{\rm uni}$ can
therefore be obtained.
 
\begin{figure}
\begin{center}
\includegraphics*[width=0.75\columnwidth]{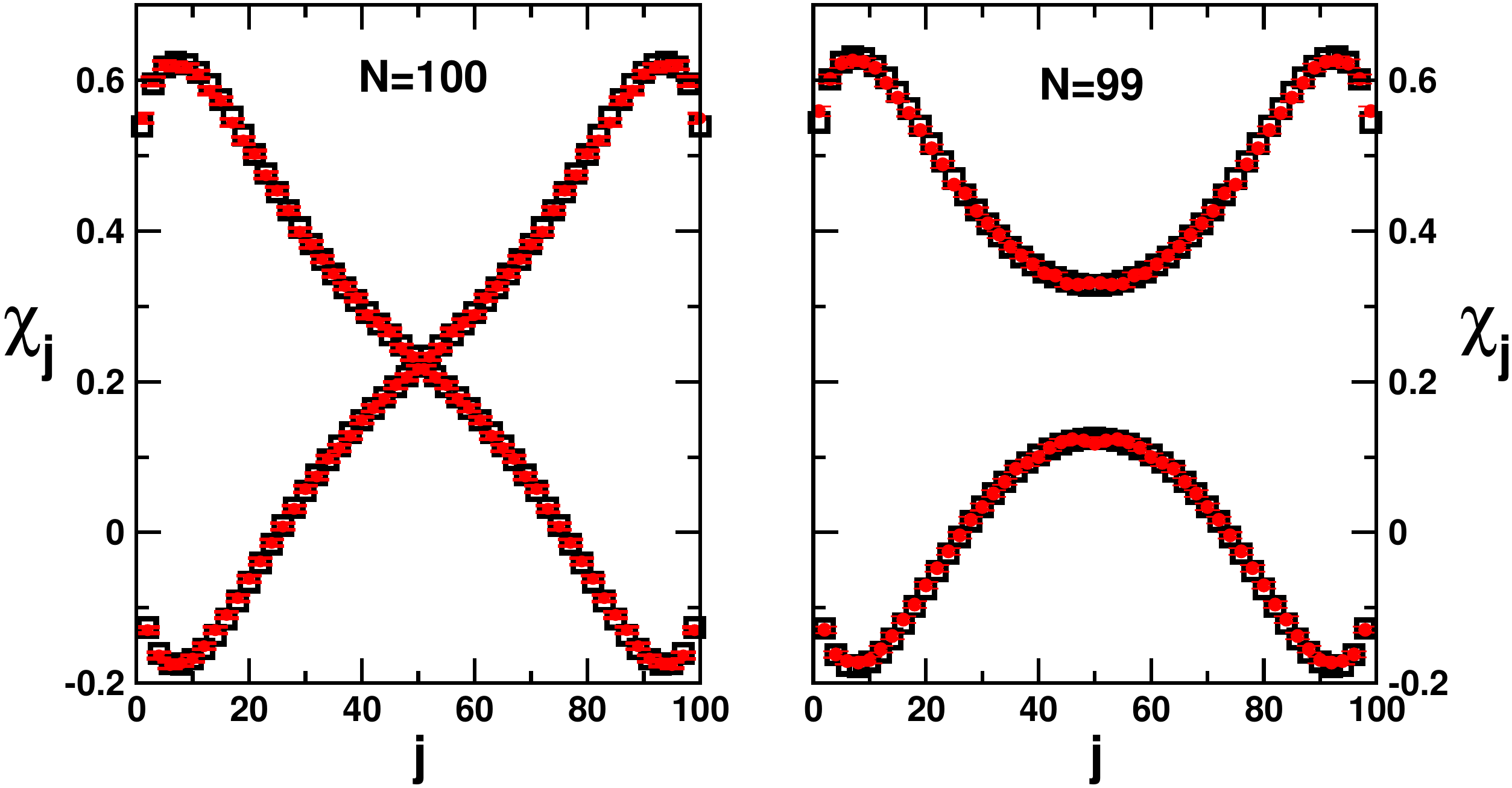}
\end{center}
\caption{Local susceptibility for a finite open $XXZ$ chain with
  length $N$, $\Delta=0.3$ at temperature $T/J=0.02$. The squares denote the result of the parameter-free field theory formula, the circles are results obtained by Quantum Monte Carlo calculations.\cite{SirkerLaflorencieEPL}}
\label{chij_fig}
\end{figure}

The alternating part of the susceptibility (\ref{eq2}) can be written
as $ \chi_j^{\rm st} = \frac{c}{T}\exp(-2\pi
K\langle\phi\phi\rangle_{\rm osc})\langle\cos\sqrt{4\pi
  K}\phi \rangle_{S_z}$ where we have again split the
correlation function into an oscillator and a zero mode part using the
mode expansion. The calculation is now completely analogous to the
calculation of the correction (\ref{free_E_corr}) leading to
\begin{equation}
\label{StaggPart2}
\chi_j^{\rm st}=-\l(\frac{\pi}{N+1}\r)^{K}\!\!\!\!\!\!\frac{\eta^{3K}\l(\e^{-\frac{\pi
    v}{TL}}\r)}{\theta_1^{K}\l(\frac{\pi j}{N+1},\e^{-\frac{\pi
    v}{2TL}}\r)} \frac{\sum_{m} m\sin[2\pi m j/(N+1)]\e^{-\pi vm^2/(2KLT)}}{\sum_{m}\e^{-\pi
    vm^2/(2KLT)}} .
\end{equation}
Here $\eta(q)$ is the Dedekind eta-function, and $\theta_1(u,q)$ the
elliptic theta function of the first kind. In the thermodynamic limit,
$L=Na\to\infty$, where we have reintroduced the lattice constant $a$
for clarity, we can simplify our result and obtain
\begin{equation}
\label{StaggPart4}
\chi_j^{\rm st} = \frac{2cK}{v}\frac{x}{\l[\frac{v}{\pi T}\sinh\l(\frac{2\pi T x}{v}\r)\r]^{K}}
\end{equation}
with $x=ja$. This agrees for the isotropic Heisenberg case, $K=1/2$,
with the result in Ref.~\citeonline{EggertAffleck95}. In
Fig.~\ref{chij_fig} the parameter-free formula for $\chi_j=\chi^{\rm
  uni}+(-1)^j\chi_j^{\rm st}$ is compared to Quantum-Monte-Carlo
data.\cite{SirkerLaflorencieEPL}

The position dependent susceptibility is directly measured as Knight
shift in NMR. The hyperfine interaction couples nuclear and electron
spins and the Knight shift of the nuclear resonance frequency for a
chain segment of length $N$ is given by
$K^{(N)}_j=(\gamma_e/\gamma_n)\sum_{j'}A^{j-j'}\chi^{(N)}_{j'}$, where
$\gamma_e$ ($\gamma_n$) is the electron (nuclear) gyromagnetic ratio,
respectively. The hyperfine interaction is short ranged so that
usually only $A^0$ and $A^{\pm 1}$ matter. For a random distribution
of non-magnetic impurities within a chain the NMR spectrum reflects
the distribution of Knight shifts for an ensemble of spin chains with
random lengths. This leads to rather complicated NMR
spectra\cite{TakigawaMotoyama,BoucherTakigawa} whose properties can be
fully understood using the parameter-free results for the
susceptibility discussed above.\cite{SirkerLaflorencieEPL}


\subsection{The spin-lattice relaxation rate and transport}
\label{Transport}
The spin current (or particle current in the fermionic language) for
the $XXZ$ model is defined by \be
\mathcal{J}=-\frac{iJ}{2}\sum_l\l(S^+_lS^-_{l+1}-S^+_{l+1}S^-_l\r)\approx
-\sqrt{\frac{K}{\pi}}\int dx\,\partial_x\theta \ee Whether or not the
integrable $XXZ$ model supports ballistic transport at finite
temperatures has been the topic of a long-standing
debate.\cite{ZotosPrelovsek,AlvarezGros,Zotos,HeidrichMeisner,SirkerPereira,SirkerPereira2}
As discussed in Sec.~\ref{Intro_Transport} ballistic transport is
signalled by a non-zero Drude weight and related to the part of the
current which cannot decay due to conservation laws, see
Eq.~(\ref{Drude}). For finite magnetic field, the Mazur inequality
indeed immediately yields a non-zero Drude
weight.\cite{ZotosPrelovsek} In this case the conserved energy current
(\ref{JE_cont}) becomes \be
\label{conserved_finite_field}
\tilde{\mathcal{J}}^E_0=-v^2\int dx \, \partial_x\theta
\,\partial_x\phi-hv\sqrt{\frac{K}{\pi}}\int dx \,
\partial_x\theta=\mathcal{J}^E_0+h\mathcal{J} \ee where we have used
again the shift in the boson field, Eq.~(\ref{Shift}). The equal time
correlations in (\ref{Drude}) can now be evaluated for the Luttinger
model (\ref{eq:Hluttingerrescaled}) and\cite{SirkerPereira2}
\begin{equation}
D\geq D_{\rm Mazur}=\frac{1}{2TL}\frac{\langle
    \mathcal{J}\tilde{\mathcal{J}}^{E}_0\rangle^{2}}{\langle (\tilde{\mathcal{J}}^{E}_0)^{2}\rangle}=\frac{vK/2\pi}{1+\frac{\pi^{2}}{3K}\left(\frac{T}{h}\right)^{2}} \qquad (T,h\ll J).
\label{eq:mazurbound}
\end{equation}
For $T/h\to 0$ the Mazur bound obtained from the overlap with
$\tilde{\mathcal{J}^E_0}$ saturates the exact zero temperature Drude
weight $D(T=0)=vK/2\pi$ \cite{ShastrySutherland}. Furthermore, one can
also use the Bethe ansatz to calculate the Mazur bound $D_{\rm Mazur}$
exactly. The obtained result agrees with (\ref{eq:mazurbound}) up to
temperatures of order $J$.\cite{SirkerPereira2}

For zero magnetic field, however, the overlap between all local
conserved quantities $\mathcal{Q}_n$ of the $XXZ$ model which can be
constructed from the transfer matrix (\ref{transfer_matrix}) and the
current $\mathcal{J}$ vanishes, because the $\mathcal{Q}_n$ are even
under particle-hole transformations while $\mathcal{J}$ is odd.
Recently, a quantity---not related to the conserved quantities
obtained from the Bethe ansatz solution---has been constructed for an
open $XXZ$ chain which is conserved up to boundary terms.\cite{Prosen}
This quantity seems to protect part of the current in the
thermodynamic limit, a view which appears to be supported by new
numerical data.\cite{KarraschMoore} As in the finite field case, the
correct picture therefore seems to be that at finite temperatures a
diffusive and a ballistic transport channel
coexist.\cite{SirkerPereira} For temperatures $T/J\in [0.2,0.5]$ and
$h=0$, the Drude weight, however, seems to be much suppressed compared
to its zero temperature value known exactly from Bethe ansatz.
Furthermore, the Drude weight seems to vanish completely at finite
temperatures in the isotropic case, $\Delta=1$. We therefore ignore a
protected part of the current for now and first concentrate on the
diffusive channel. The corrections due to a possible conserved part of
the current are discussed at the end of this section.

\begin{figure}
\begin{center}
\includegraphics*[width=0.75\columnwidth]{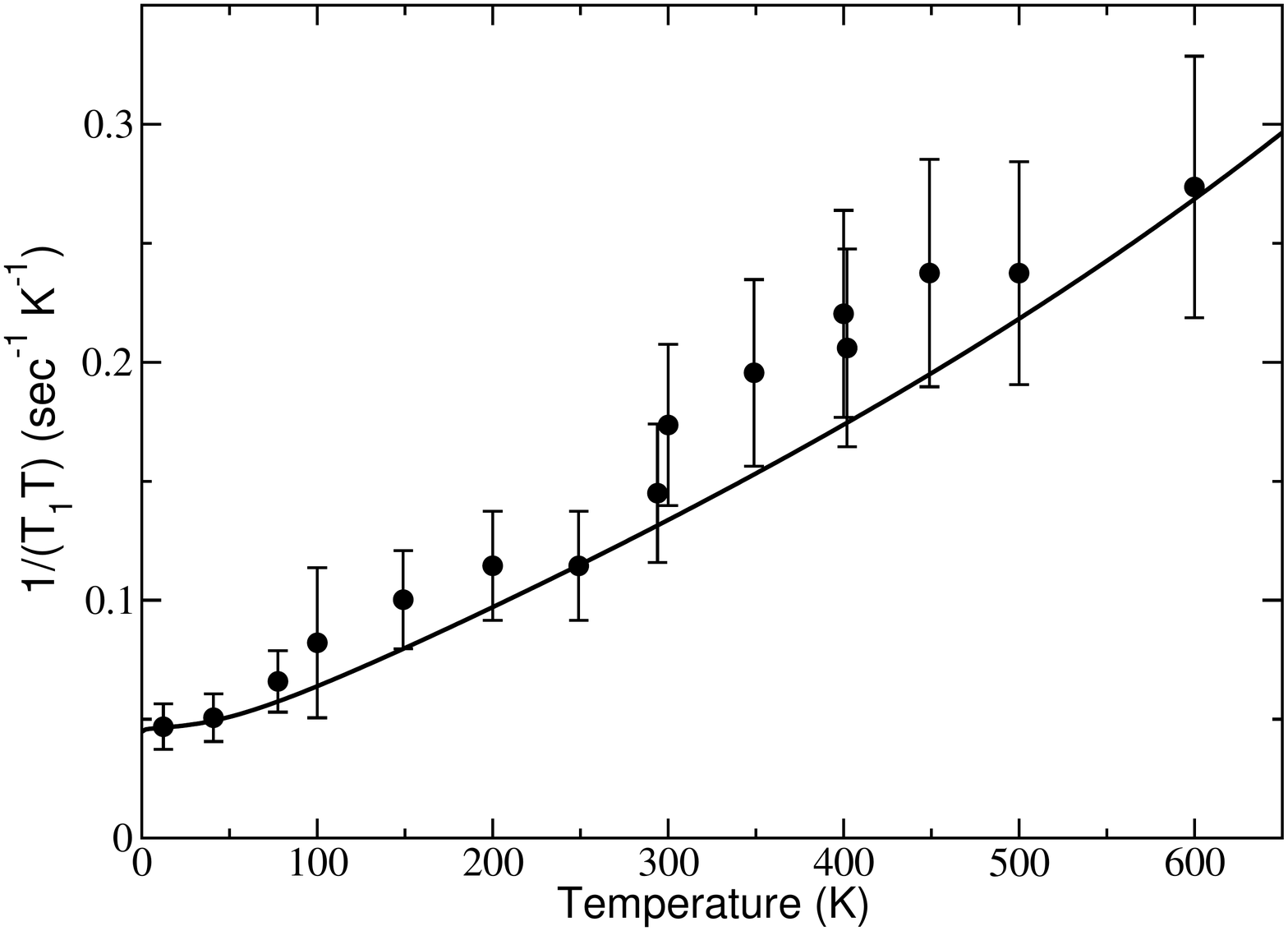}
\end{center}
\caption{Spin-lattice relaxation rate $1/T_1$ for Sr$_2$CuO$_3$.
  Experimental data (symbols) are compared to the field theoretical
  result. The only free parameters are the exchange constant $J\sim
  1800$ K and the hyperfine coupling tensor which have been determined
  experimentally. A more detailed analysis is presented in
  Ref.~\citeonline{SirkerPereira2}.}
\label{T1_fig}
\end{figure}

From an experimental point of view, it is of great interest to
understand the intrinsic mechanism which leads to spin diffusion in
the $XXZ$ model. Spin diffusion has directly been observed in the spin
lattice relaxation rate $1/T_1$ measured in NMR experiments as a
magnetic field dependence $1/T_1\sim
1/\sqrt{h}$.\cite{ThurberHunt,PrattBlundell} Both the spin lattice
relaxation rate and the conductivity can be calculated within the
Luttinger model from the retarded boson propagator
\begin{equation}
\label{SelfE}
\langle\phi\phi\rangle^{\rm ret}(q,\omega)=\frac{v}{\omega^2-v^2q^2-\Pi^{\rm ret}(q,\omega)}.
\end{equation}
For $\Pi^{\rm ret}(q,\omega)=0$ this is just the free boson
propagator. In the zero field case, the leading irrelevant operator is
the Umklapp term (\ref{H_U}) and we will concentrate here on
calculating the self-energy $\Pi^{\rm ret}(q,\omega)$ in first order
in this perturbation. Further contributions to the self-energy will
stem from the band curvature terms and are discussed in
Ref.~\citeonline{SirkerPereira2}. Note, however, that only Umklapp
scattering can give the bosons a finite lifetime and thus lead to
diffusive transport. The calculation of the correlation function
(\ref{SelfE}) in first order in Umklapp scattering is
straightforward.\cite{OshikawaAffleck02} We
find\cite{SirkerPereira,SirkerPereira2} \be \Pi^{\rm ret}(q,\omega)
\approx-2i\gamma\omega
\label{Pi2ndorder} 
\ee 
where the decay rate $\gamma$ is given by
\begin{equation}
\label{selfE_parameters}
2\gamma = 8\pi K \lambda^2\sin(4\pi
K)\l(\frac{2\pi}{v}\r)^{8K-2}\!\!\!\!\!\!\!\Gamma(1/2-2K)\Gamma(2K) \frac{B(2K,1-4K)}{\sqrt{\pi}2^{4K+1}}\cot(2\pi K) T^{8K-3}
\end{equation}
in the anisotropic case and by
\begin{equation}
\label{selfE_parameters_iso}
2\gamma = \pi g^2T
\end{equation}
in the isotropic case, $\Delta=1$. The running coupling constant
$g=g(T)$ is determined by (\ref{running_r}). The Kubo formula directly
relates the conductivity $\sigma(q,\omega)$ to the calculated bosonic
Green's function
\begin{equation}
\label{Kubo2}
\sigma(q,\omega) = \frac{K}{\pi}\im\omega \langle
\phi\phi\rangle^{ret}(q,\omega) 
\end{equation} 
and due to the finite relaxation rate $\gamma$ at finite temperatures
one find a Lorentzian for the real part of the conductivity \be
\sigma^\prime(q=0,\omega)=\frac{vK}{\pi}\frac{2\gamma}{\omega^2+(2\gamma)^2}.
\ee At the same time, also the spin lattice relaxation rate can be
expressed by the same bosonic Green's function \be
\frac{1}{T_1}\approx-\frac{2KT}{\pi\omega_e}\int\frac{q^2dq}{2\pi}\,|A(q)|^2\,\textrm{Im
}\langle\phi\phi\rangle_{\rm ret}(q,\omega_e).
\label{1T1formula}
\ee Here $\omega_e=\mu_B h$ is the electron magnetic resonance
frequency.\cite{SirkerPereira2} If the hyperfine coupling form factor
$A(q)$ picks out the $q\sim 0$ contributions of the integral
(\ref{1T1formula}) as in the oxygen NMR experiment on Sr$_2$CuO$_3$ in
Ref.~\citeonline{ThurberHunt} then $1/T_1$ and the conductivity
$\sigma'(q\sim 0,\omega)$ are directly related. The experimentally
considered spin chains are almost isotropic, $\Delta\approx 1$. Using
the parameter-free result (\ref{selfE_parameters_iso}) we obtain the
diffusive behavior \be \frac{1}{T_1T}\sim
\sqrt{\frac{\gamma(T)}{\omega_e}}\sim
\sqrt{\frac{T/\ln^2(J/T)}{\omega_e}}.
\label{1T1short}
\ee The only free parameters remaining depend on microscopic details
of the considered compound. Both the exchange constant $J$ and the
hyperfine coupling constant $A(q)$ can be fixed by analyzing the
susceptibility and performing a $K-\chi$ analysis
respectively.\cite{MotoyamaEisaki,ThurberHunt,BoucherTakigawa}
Integrability therefore makes it possible to obtain an analytical
result for the spin-lattice relaxation rate which includes the
intrinsic relaxation processes. The comparison of the result
(\ref{1T1short}) with experiment as shown in Fig.~\ref{T1_fig}
demonstrates that Umklapp scattering seems to be the dominant source
for relaxation and that other contributions, e.g., due to
electron-phonon scattering are small.

Finally, we have to discuss the possible conservation of part of the
current even at zero magnetic field. The quantity constructed in
Ref.~\citeonline{Prosen} contains terms active over several lattice
sites for $0<\Delta<1$. It is therefore non-trivial to bosonize this
term and to consider the consequences of its (almost) conservation for
the low-energy effective theory in a similar spirit to our discussion
of $J^E$ in Sec.~\ref{coup_const}. An alternative method discussed in
Refs.~\citeonline{RoschAndrei,SirkerPereira,SirkerPereira2} is to use
a memory matrix formalism to calculate the self-energy (\ref{SelfE})
in the presence of conservation laws. Instead of
Eq.~(\ref{Pi2ndorder}) the self-energy then reads \be
\Pi(q=0,\omega)\approx \frac{-2i\gamma\omega}{1+2i\gamma y/\omega}
\quad \mbox{with}\quad y=\frac{\langle \mathcal{J} Q\rangle^2}{\langle
  \mathcal{J}^2\rangle \langle Q^2\rangle -\langle \mathcal{J}
  Q\rangle^2 }
\label{mem_matrix}
\ee where $Q$ is the conserved quantity with
$\langle \mathcal{J}Q\rangle\neq 0$. As a consequence, the
current-current correlation function at large times is now given by \be
\frac{1}{L}\langle \mathcal{J}(t)\mathcal{J}\rangle
=\frac{KvT}{\pi(1+y)}\left[y+\exp(-2\gamma(1+y)t)\right] \ee
where the first term is proportional to the Drude weight and the
second term describes the diffusive part. 
For $\Delta=1$ the Drude weight seems to be zero at finite
temperatures\cite{Prosen} so that our theoretical analysis of $1/T_1$
is not affected. In any case, the spin chains in Sr$_2$CuO$_3$ do not
represent an integrable system and there is therefore certainly no
ballistic channel. It is known, for example, that in this compound
also a weak next-nearest neighbor coupling $J_2\sim 0.1 J$ exists.
While this coupling destroys integrability it will only lead to a weak
renormalization of the Umklapp amplitude
$\lambda$.\cite{EggertAffleck92} It is thus legitimate to use the
results for the integrable model to analyze the experimental data for
$1/T_1$.

\section{Conclusions}
\label{Concl}
Integrable gapless one-dimensional quantum models allow to test many
predictions of Luttinger liquid theory. As such they have been vital
in confirming the universal applicability of the latter. On the other
hand, Luttinger liquid theory has also helped to understand integrable
quantum models better. Except for the simplest integrable systems,
such as free bosonic or fermionic particles and to some extent also
Calogero-Sutherland type models, an exact calculation of correlation
functions for arbitrary distances and times has not been achieved yet.
A combination of Luttinger liquid theory and integrability then often
allows to obtain a more complete picture. For Bethe ansatz integrable
systems such as the Lieb-Liniger Bose gas, the (anisotropic)
Heisenberg, the Hubbard, or the supersymmetric $t-J$ model it is
possible to fix the velocity of the collective excitations $v$ and the
Luttinger parameter $K$ as a function of density and interaction
strength. For the Luttinger model, (dynamical) correlations can then
be easily calculated. 

Often one can go even one step further and determine the amplitudes of
leading irrelevant operators acting as corrections to the Luttinger
Hamiltonian as well as amplitudes of correlation functions, exactly.
This article is by no means a complete review of all the results which
have been obtained in this field. Instead, I have used the $XXZ$ model
as an example and have summarized how $v$, $K$ as well as the
amplitudes of the leading irrelevant band curvature terms and Umklapp
scattering can be determined. As an application, I have shown that
this allows to derive parameter-free results for the local
susceptibility in open Heisenberg chains and quantitatively explains
Knight shift spectra which have been investigated by nuclear magnetic
resonance for compounds such as Sr$_2$CuO$_3$. As a second
application, I have presented results for the relaxation rate of the
Luttinger liquid bosons at finite temperatures in first order in
Umklapp scattering.  From this a parameter-free formula for the
spin-lattice relaxation rate $1/T_1$ can be obtained which is in
excellent agreement with experiment. The same bosonic correlation
function and the same relaxation rate determine, on the other hand,
also the particle transport. In general, the $XXZ$ model has
coexisting ballistic and diffusive transport channels. The relative
weight of each of these channels can be determined by combining the
Luttinger model---keeping the dangerously irrelevant Umklapp term with
known amplitude---with a memory-matrix calculation.

There are many more interesting results which have not been covered in
this review. One of the perhaps most fascinating recent developments
is the so-called non-linear Luttinger liquid theory which allows to
calculate dynamic response functions while taking band curvature into
account.\cite{ImambekovGlazman} This has lead to the discovery of new
power laws near edge singularities. For integrable models the
exponents of these power laws can be determined
exactly.\cite{PereiraWhite} A comprehensive overview about these
developments has been given in a recent excellent
review.\cite{GlazmanReview}

\section*{Acknowledgements}
I acknowledge support by the excellence graduate school MAINZ
and the collaborative research center SFB/TR49.




\end{document}